\documentclass[11pt,twocolumn,twoside]{article}
\usepackage{fully3d}

\begin{document}

\title{Noise Entangled GAN For Low-Dose CT Simulation} 

\author[1]{Chuang Niu}
\author[1]{Ge Wang}
\author[1]{Pingkun Yan}
\author[1]{Juergen Hahn}
\author[2]{Youfang Lai}
\author[2]{Xun Jia}
\author[3]{Arjun Krishna}
\author[3]{Klaus Mueller}
\author[4]{Andreu Badal}
\author[4]{Kyle J. Myers}
\author[4]{Rongping Zeng}

\affil[1]{Department of Biomedical Engineering, Center for Biotechnology \& Interdisciplinary Studies, Rensselaer Polytechnic Institute, Troy, NY USA}

\affil[2]{Department of Radiation Oncology, UT Southwestern Medical Center, Dallas, TX USA}
          
\affil[3]{Computer Science Department, Stony Brook University, Stony Brook, NY USA}

\affil[4]{Division of Imaging, Diagnostics and Software Reliability, OSEL, CDRH, U.S. Food and Drug Administration, Silver Spring, MD USA}

\maketitle
\thispagestyle{fancy}


\begin{customabstract}
We propose a Noise Entangled GAN (NE-GAN) for simulating low-dose computed tomography (CT) images from a higher dose CT image. First, we present two schemes to generate a clean CT image and a noise image from the high-dose CT image. Then, given these generated images, an NE-GAN is proposed to simulate different levels of low-dose CT images, where the level of generated noise can be continuously controlled by a noise factor. NE-GAN consists of a generator and a set of discriminators, and the number of discriminators is determined by the number of noise levels during training. Compared with the traditional methods based on the projection data that are usually unavailable in real applications, NE-GAN can directly learn from the real and/or simulated CT images and may create low-dose CT images quickly without the need of raw data or other proprietary CT scanner information. The experimental results show that the proposed method has the potential to simulate the realistic low-dose CT images.
\end{customabstract}



\section{Introduction}

An excess of x-ray exposure from computed tomography (CT) examinations could lead to the development of cancer, and thus optimizing CT protocols according to the as low as reasonably achievable (ALARA) principle has become important. Low-dose CT (LDCT) simulation techniques have developed as an effective tool to help determine the lowest dose in accordance with the ALARA principle, thereby circumventing the repetition of CT examinations with different exposure conditions for the same patients. However, reducing the radiation dose will inevitably increase the noise level in the reconstructed CT images and may compromise the accuracy of a radiologist's diagnostic decision. To this end, a lot of LDCT denoising methods have been proposed to improve the image quality. Recently, deep-learning-based denoising methods have been shown a potential to achieve the superior denoising performance, if properly trained with a large number of CT images. In this context, the results with LDCT simulation methods can help train and test the robustness of denoising methods or other image analysis models applied to the LDCT images.

Traditionally, LDCT simulation tools insert random noise to the raw sinogram data and reconstruct the noisy data to simulate LDCT images \cite{c11}. However, neither raw data nor the precise parameters of a CT imaging system are generally accessible without an established collaboration with the CT vendor. To circumvent the use of raw data, projection data can be approximated by forward projecting from the CT image, which are then added with noise and reconstructed using CT simulation software \cite{c16}. However, these sinogram-based methods are usually time-consuming and the simulated projection data may not truly reflect the real conditions so the simulated LDCT noise is likely still not perfect. Recently, Shan et al. designed a specific GAN with a conditional batch normalization layer to simulate LDCT noise from a random 2-dimensional Gaussinan noise vector in the latent space \cite{c19}. However, it is difficult for this method to generate realistic LDCT images from the Gaussian noise without explicit prior information of the LDCT noise. 

In this work, we treat the LDCT simulation as a transformation from a higher dose CT (HDCT) image to the LDCT images. Specifically, we first generate a clean CT image and a high-dose noise image from the HDCT image, and then train a noise entangled GAN (NE-GAN) to generate different levels of LDCT images via entangling the high-dose noise image scaled by different noise factors into the clean CT image. 
The advantages of the proposed framework for LDCT simulation are: 1) The generated high-dose noise image explicitly contains the prior of noise and imaging system. 2) The NE-GAN can learn from both the simulated and real CT images, so that it has the potential to generate realistic LDCT images. 3) Once the model has been trained, the simulation speed for LDCT images is very fast.


\section{Methods}




Ideally, before simulating the LDCT image from the HDCT image, the noise component should be removed from the HDCT image and then low-dose noises are simulated and added to the denoised image. In practice, the HDCT images are usually regarded as the clean image and the high-dose noises are ignored. Although the magnitudes of high-dose noises are low, they do contain the prior information of CT noises and the imaging system to some extent. Based on above observations, we propose to simulate an LDCT image through two steps: the fist step is to generate a clean image and a high-dose noise image from the HDCT image, and the second step is to generate different levels of LDCT images by entangling the high-dose noise component scaled with a specific noise factor into the clean CT image.

\begin{figure}[hbt!]
    \centering
    \includegraphics[width=0.40\textwidth]{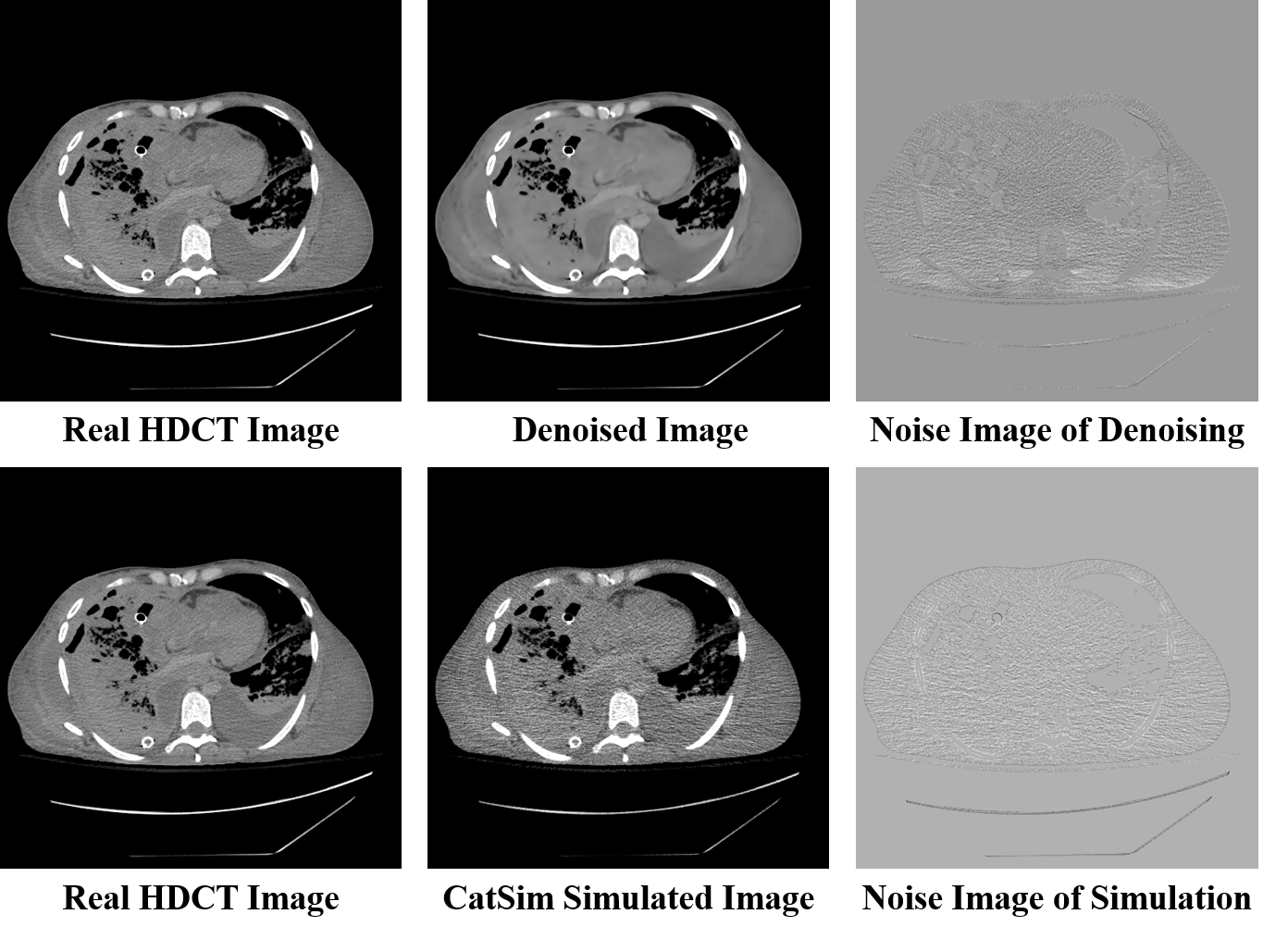}
    \caption{Generation of noise image.}
    \label{fig_ng}
\end{figure}

\subsection{Generation of high-dose noise image}
\label{sec_gn}

\begin{figure}[hbt!]
    \centering
    \includegraphics[width=0.49\textwidth]{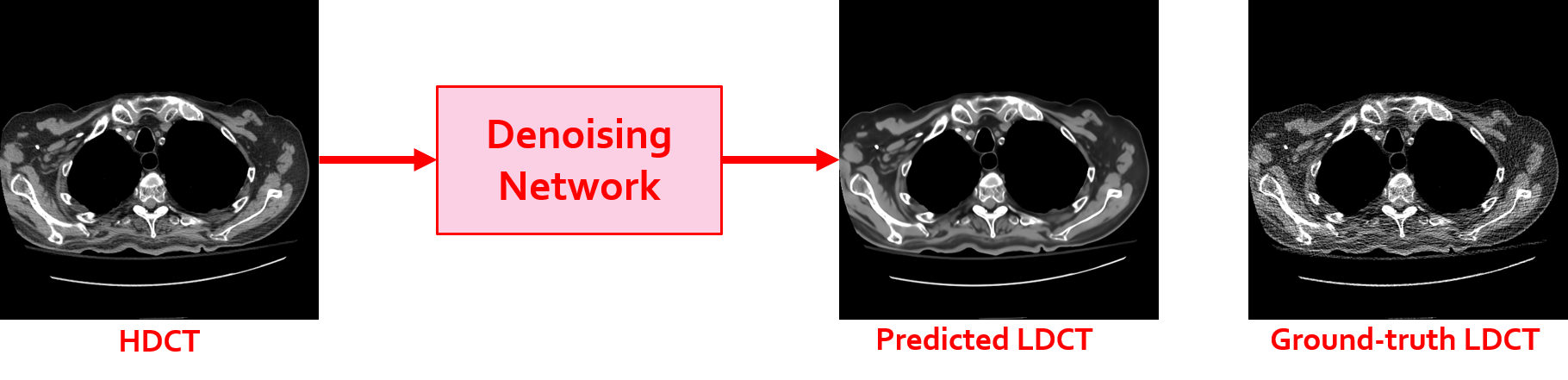}
    \caption{A denoising network trained using HDCT as input and LDCT as target. The predicted "LDCT" is a cleaner CT image rather than a noisier.}
    \label{fig_denoising}
\end{figure}

For generating the high-dose noise image that preserves the prior information of the CT noise and imaging system, we present two schemes, i.e., through a denoising network or through CT simulation, as illustrated in Fig. 1. For the denoising scheme, the HDCT image is first forwarded into a denoising model to obtain a clean CT image, which is then subtracted from the input HDCT image to generate the high-dose noise image.
The denoising model is trained by directly mapping high-dose CT image to the low-dose CT image, as shown in Fig. \ref{fig_denoising} . By doing this, the trained model can generate the denoised image instead of the images with more noises, which is consistent to the findings of Noise2Noise \cite{c20}. The denoising scheme can extract the real prior information from the real HDCT images, which are then transformed to LDCT images with specific noise level by NE-GAN.
The CT simulation scheme is to use traditional sinogram-based methods to simulate a set of higher-dose noise images by virtually scanning the real HDCT image, and the real HDCT image is regarded as the clean CT image, as shown in Fig. \ref{fig_ng}. Then, NE-GAN takes the simulated noise image and the HDCT image as inputs to generate a set of LDCT images with different levels of noise. In this scheme, the sinogram-based method is only used to simulate a single dose of images, and other lower dose of images can be generated by NE-GAN to save computation time.

\begin{figure}[hbt!]
    \centering
    \includegraphics[width=0.49\textwidth]{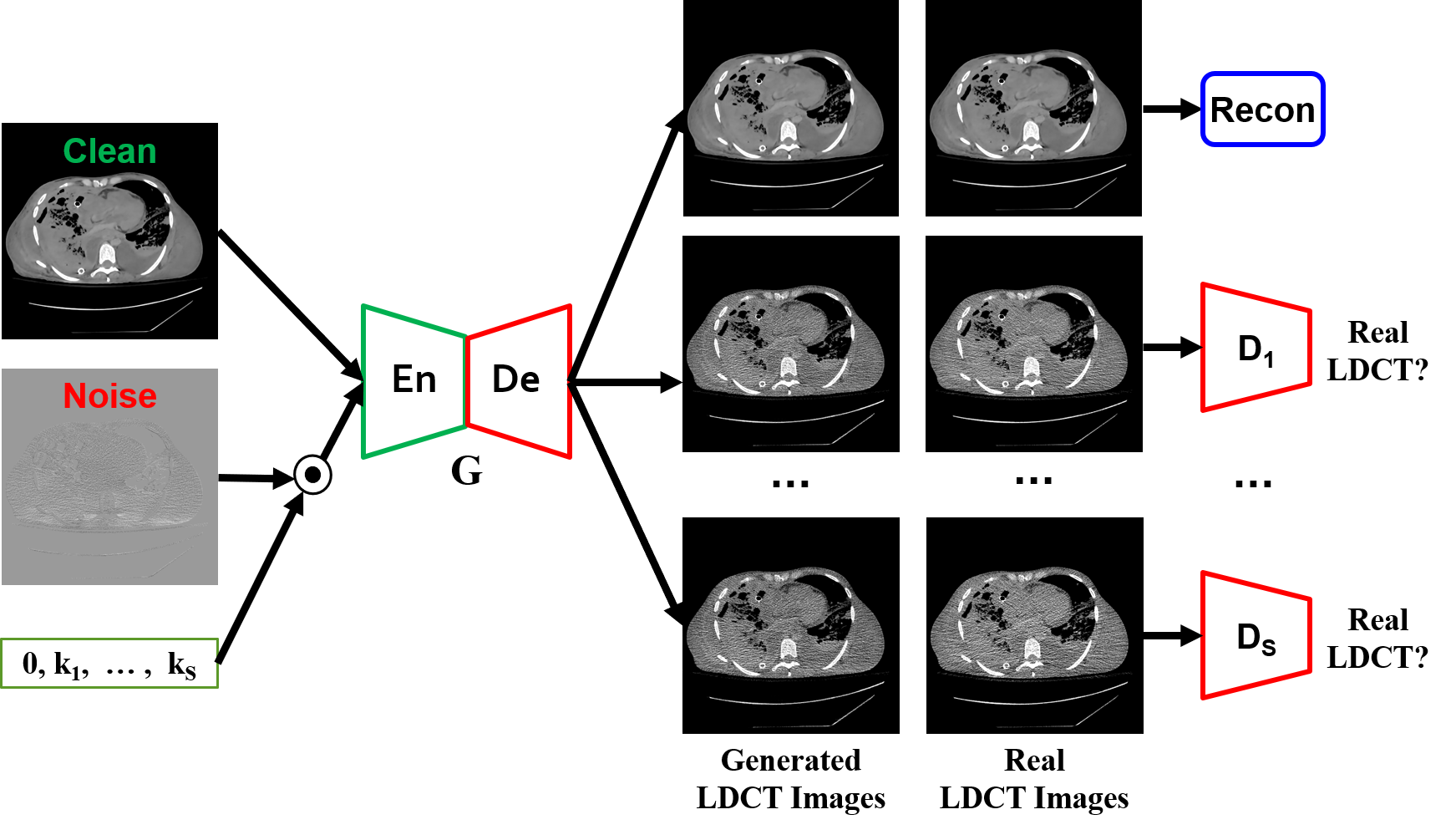}
    \caption{Framework of NE-GAN.}
    \label{fig_negan}
\end{figure}

\subsection{Noise Entangled GAN}
\label{sec_negan}

In this Subsection, we describe the details of the proposed noise entangled GAN (NE-GAN). As shown in Fig. \ref{fig_negan}, NE-GAN consists of a generator $G$ and a set of discriminators $\textbf{D} = \{ D_j \}, j =1, \cdots, S$, where the number of discriminators $S$ is equivalent to the number of lower-dose levels in the training set. Specifically, the generator $G$ is a encoder-decoder network that takes a clean CT image and a noise image scaled with a noise factor as inputs, and outputs a LDCT image corresponding to the input noise factor.
All discriminators share the same network architecture. Each discriminator is to determine whether the generated image  of a predefined level is real.

To train NE-GAN, we need a set of training samples $\{x_i^0, n_i^0, k^j, x_i^j\}$, $i=1, \cdots, N, j=1, \cdots, S$, where $x_i^0$ and $n_i^0$ denote the clean CT image and high-dose image respectively, $N$ is the total number of HDCT images, $x_i^j$ denotes the corresponding LDCT image, and $S$ is number of noise levels or discriminators, $k^j$ denotes the noise factor that is a positive real number and a larger value corresponds to a higher noise level or lower image quality. The loss function is:
\begin{align*}
        L = &\sum_{j=1}^{S} E_{X^j} [ \log D_j(x^j)] + E_{X^0} \left[log(1 - D_j(G(x^0, n^0 \cdot k_j)))\right.\\
        &\left.+| x^0 - G(x^0, n^0 \cdot k_j)  | + |x^0 - G(x^0, \emph{0})|\right].
\end{align*}
The first two items in the loss function are the adversarial losses that train $\textbf{D}$ to maximize the probability of assigning the correct label to both real LDCT images and the generated ones from $G$ and train G to minimize probability of assigning the correct label for $\emph{D}$, the third item is a data fidelity loss to constrain the generated LDCT images to keep the same contents as those in the input images, and the fourth item is a reconstruction loss to ensure that the generated CT images are exactly the clean images when the noise factor is zero.

After training, only the generator $G$ is retained to simulate different levels of LDCT images given the clean CT image, the high-dose noise image, and the specific noise factor, i.e., $\hat{x}^j = G(x^0, n^0 \cdot k_j)$. It is noted that although the noise factor in the training stage is predefined as a limited number of fixed values according to the training dataset, it could be any value in the testing stage beyond the predefined values in the training stage. With increasing the value of noise factor, the noise level of the simulated LDCT image will increase.

\subsection{Implementation details}
We adopted the same generator and discriminator networks as those in CycleGAN \cite{c18}.
The architecture of the denoising network was the same as the generator network.
During training, we used the Adam method to optimize the NE-GAN model with a batch of 8 $128 \times 128$ randomly cropped image patches. The initial learning rate was set to 0.0002 during the first 200 epochs and the learning rate was linearly decay to zero in the following 200 epochs. The momentum terms of Adam were set to 0.5 and 0.999.
The noise factor $k_j$ is set to the ratio of the input dose level to the target dose level, see Subsections \ref{sec_res_sim} and \ref{sec_res_real} for details.

\section{Experiments and results}
\label{sec_exp}

\subsection{Dataset}
In this study, we used a multi-dose of real CT image dataset from \cite{c21}, in which the CT images were collected from anonymous cadavers and each of them was repeatedly scanned four times using four different radiation doses.  
In our experiments, we selected a sub-dataset that contains 261 groups of CT images for training and 251 groups of CT images for testing, each group includes four 512 $\times$ 512 FBP reconstructed images that have the same contents but different noise indices of 10, 20, 30, and 40. Here the noise index is approximately equal to standard deviation of CT number in the central region of the image of a uniform phantom, and used to define the image quality.

\begin{figure}[hbt!]
    \centering
    \includegraphics[width=0.49\textwidth]{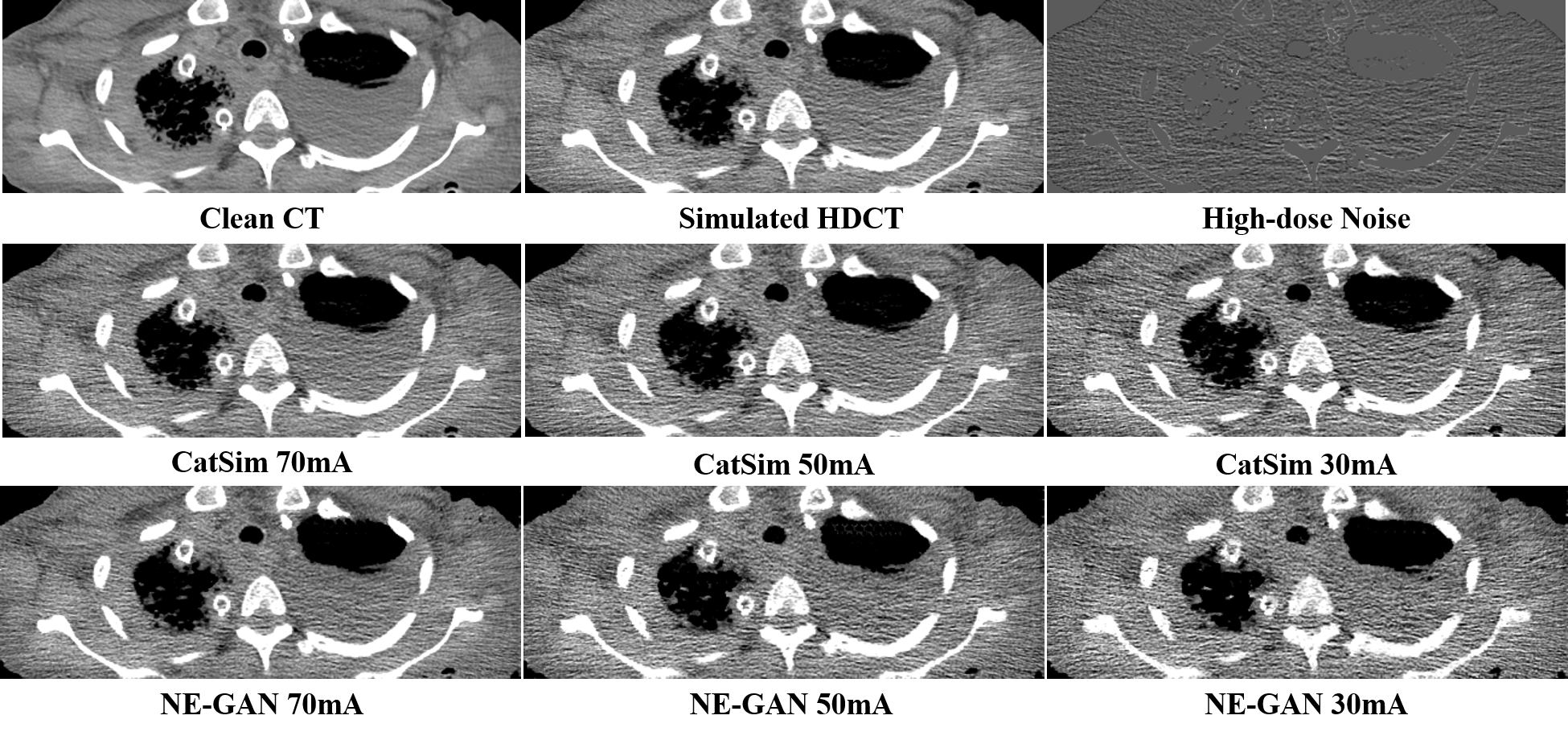}
    \caption{Results of NE-GAN on simulated dataset.}
    \label{fig_c1}
\end{figure}
\begin{figure}[hbt!]
    \centering
    \includegraphics[width=0.49\textwidth]{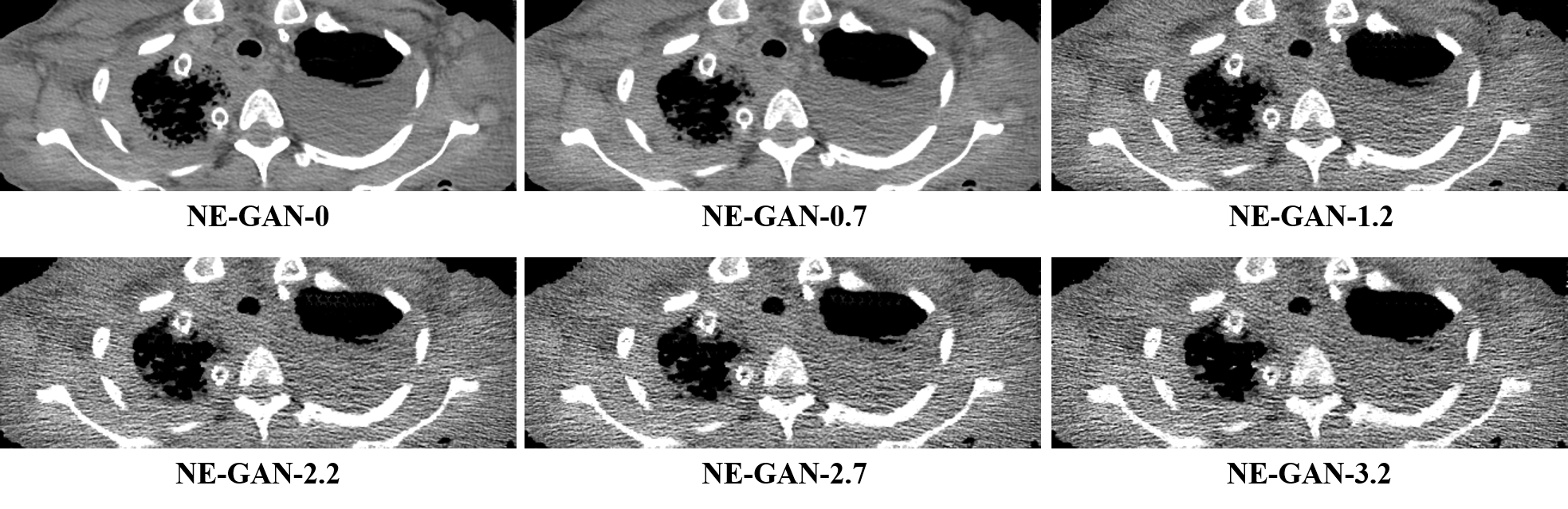}
    \caption{Results of NE-GAN on simulated dataset with additional noise factors beyond training.}
    \label{fig_c2}
\end{figure}

\subsection{Results on simulated dataset}
\label{sec_res_sim}
In this subsection, we used the simulation scheme as described in Subsection \ref{sec_gn} to generate high-dose noise images. Specifically, we used the CatSim \cite{c24} simulator to simulate LDCT images of four different dose levels corresponding to X-ray tube currents of 90 mA, 70 mA, 50 mA, and 30 mA, as shown in Fig. \ref{fig_c1}. The simulated CT image of 90 mA is used as the HDCT image and the real CT image with noise index of 10 is regarded as the clean CT image, thus the high-dose noise image is the difference between them. With these images, NE-GAN was trained and noise factors corresponding to 70 mA, 50 mA, and 30 mA were set to 1.3, 1.8, and 3.0 respectively. The results in this setting are shown in Fig. \ref{fig_c1}, we can see that the proposed method can simulate the different levels of LDCT images and the learned noise levels are similar to those simulated with CatSim. The NE-GAN simulated results with different noise factors that were not used in the training stage are shown in Fig. \ref{fig_c2}, where the number indicates the noise factor. Particularly, NE-GAN-0 means that the scale factor is zero and in this case no noises are added, consistent with the constraint in the loss function as described in Subsection \ref{sec_negan}. Also, when increasing the noise factor, the noise magnitude of simulated LDCT image increases and the image looks more noisier.

\begin{figure}[hbt!]
    \centering
    \includegraphics[width=0.49\textwidth]{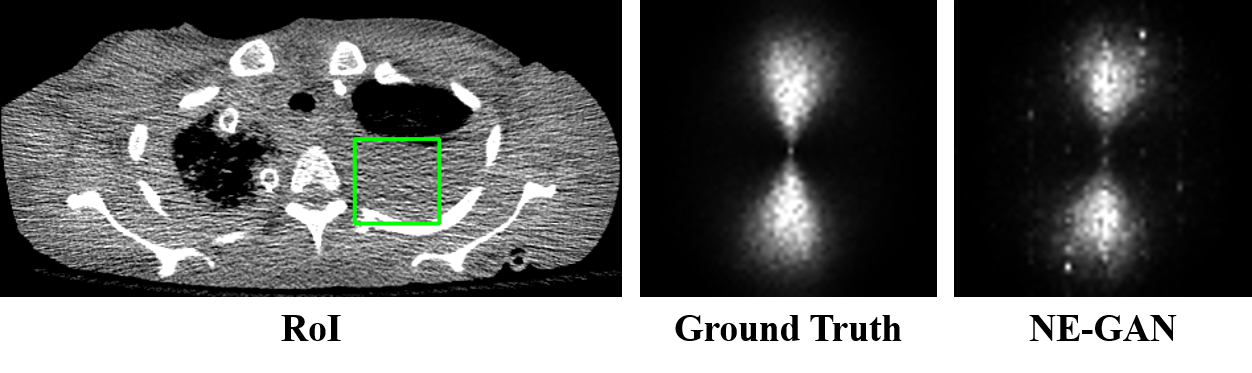}
    \caption{Results of noise power spectrum.}
    \label{fig_c3}
\end{figure}

In addition, we evaluated the statistical property of noise power spectra (NPS) of the NE-GAN generated LDCT images \cite{c23}. Specifically, we repeatly generated HDCT noise image with CatSim and simulated the LDCT images with NE-GAN by 50 times. Then, the $64 \times 64$ image patches (green box) were cropped to calcuate the NPS, as shown in Fig. \ref{fig_c3}. The NPS of the NE-GAN generated LDCT images is similar to that of the targets, which indicates the proposed deep-learning-based method has the ability to preserve the statistical properties of noise.

\begin{figure}[hbt!]
    \centering
    \includegraphics[width=0.49\textwidth]{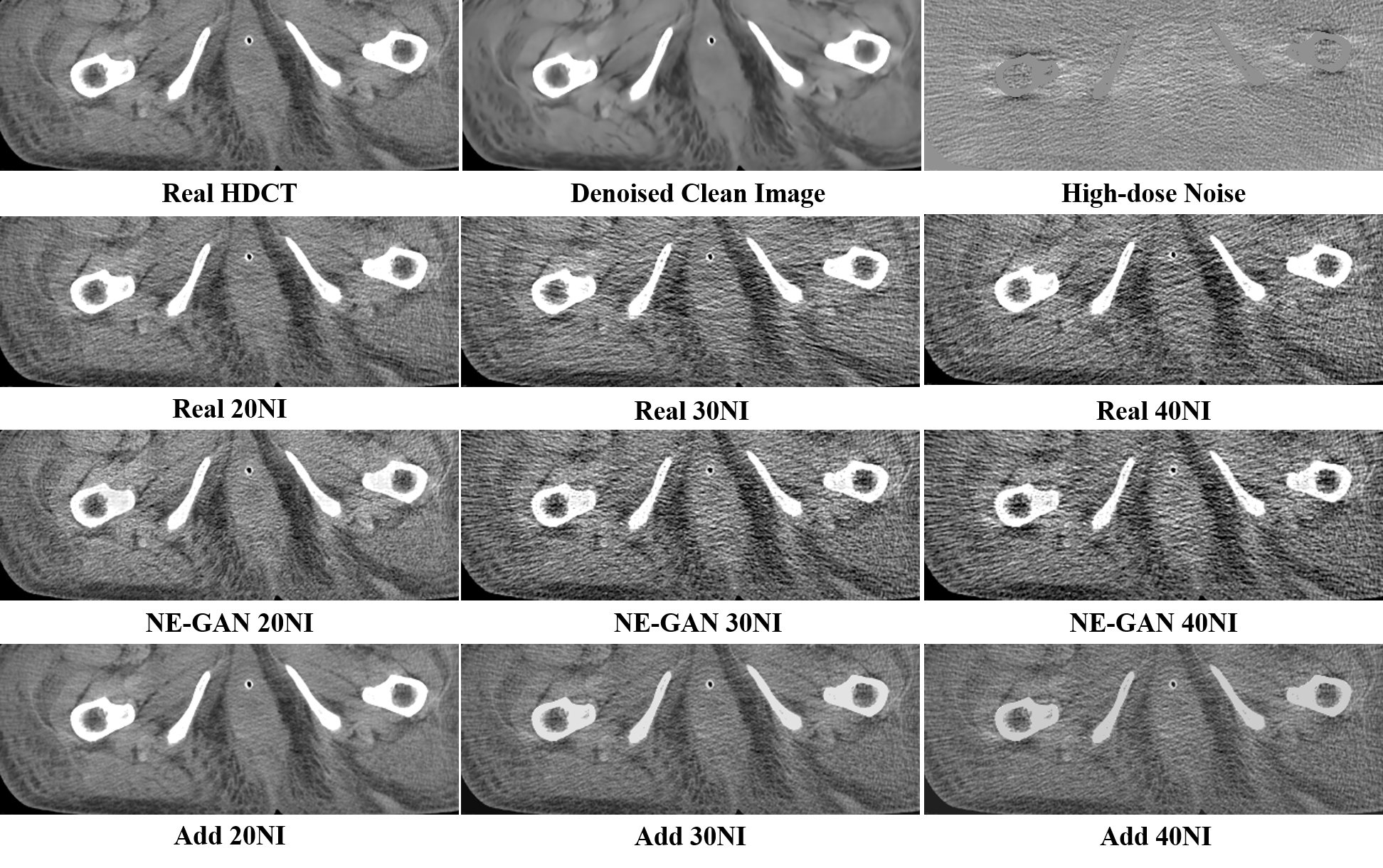}
    \caption{Results of NE-GAN on real dataset.}
    \label{fig_m1}
\end{figure}
\begin{figure}[hbt!]
    \centering
    \includegraphics[width=0.49\textwidth]{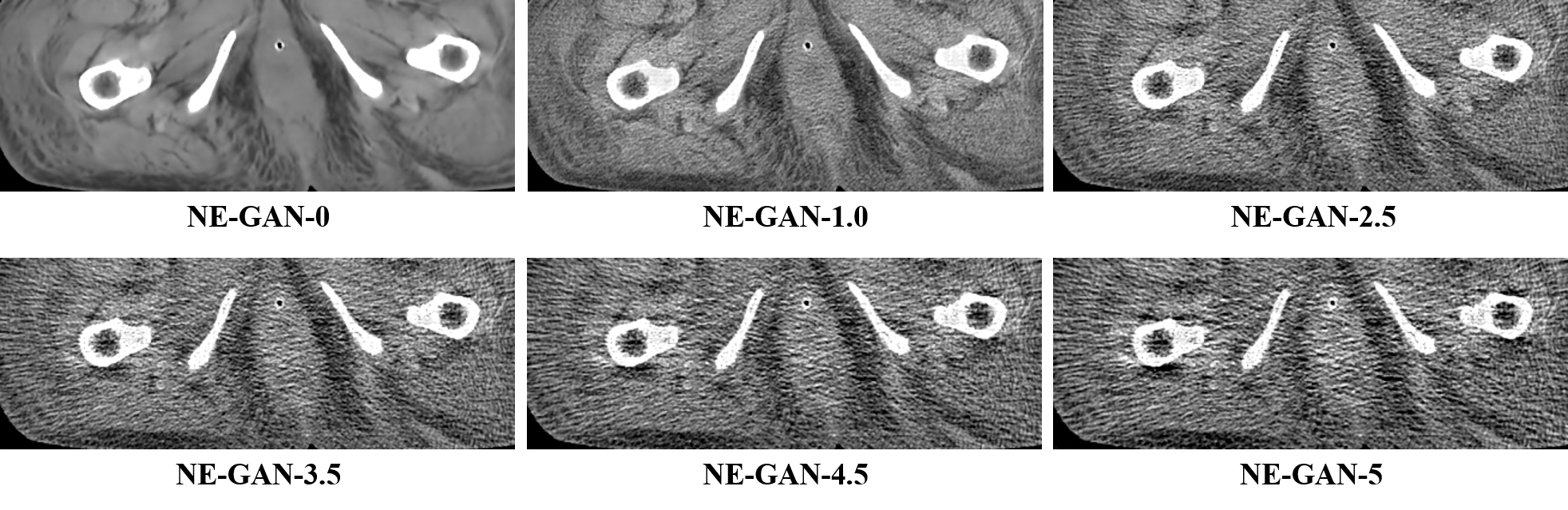}
    \caption{Results of NE-GAN on real dataset with additional noise factors beyond training.}
    \label{fig_m2}
\end{figure}

\subsection{Results on real dataset}
\label{sec_res_real}
In this subsection, the proposed NE-GAN model was directly trained on the real dataset. Here the denoising scheme was firstly used to decompose the high-dose CT image with noise index of 10 to a clean CT image and a high-dose noise image. Then the NE-GAN was trained to map these decomposed images to the LDCT images with specific noise indices. The noise factors corresponding to noise indices of 20, 30, and 40 were set to 2.0, 3.0, and 4.0 respectively. The results on this real dataset set are shown in Fig. \ref{fig_m1}, where the first row shows the denoised clean image and the high-dose noise image decomposed from the real HDCT image with noise index of 10, the second row presents the real LDCT images with different noise indices, the third row gives the corresponding NE-GAN generated LDCT images with the same noise indices, and the last row shows the reference results by directly adding the scaled noise image with the same noise factors into the clean image. By comparing the second and the third row, we can see that the simulated images of different noise indices are similar to the corresponding real LDCT images. The results in the last row demonstrates that simply scaling the extracted noise image and adding it back to the clean image cannot generate images matching with the real LDCT images, while the proposed NE-GAN has the ability to simultaneously transfer and merge the high-dose noise image into the clean image to simulate more realistic LDCT images. More simulation results with NE-GAN with different noise factors beyond training are also shown in Fig. \ref{fig_m1}. Similarly, noise level increased continuously with the noise factors.

\section{Discussion and Conclusion}
\label{sec_conclud}

\begin{figure}[hbt!]
    \centering
    \includegraphics[width=0.49\textwidth]{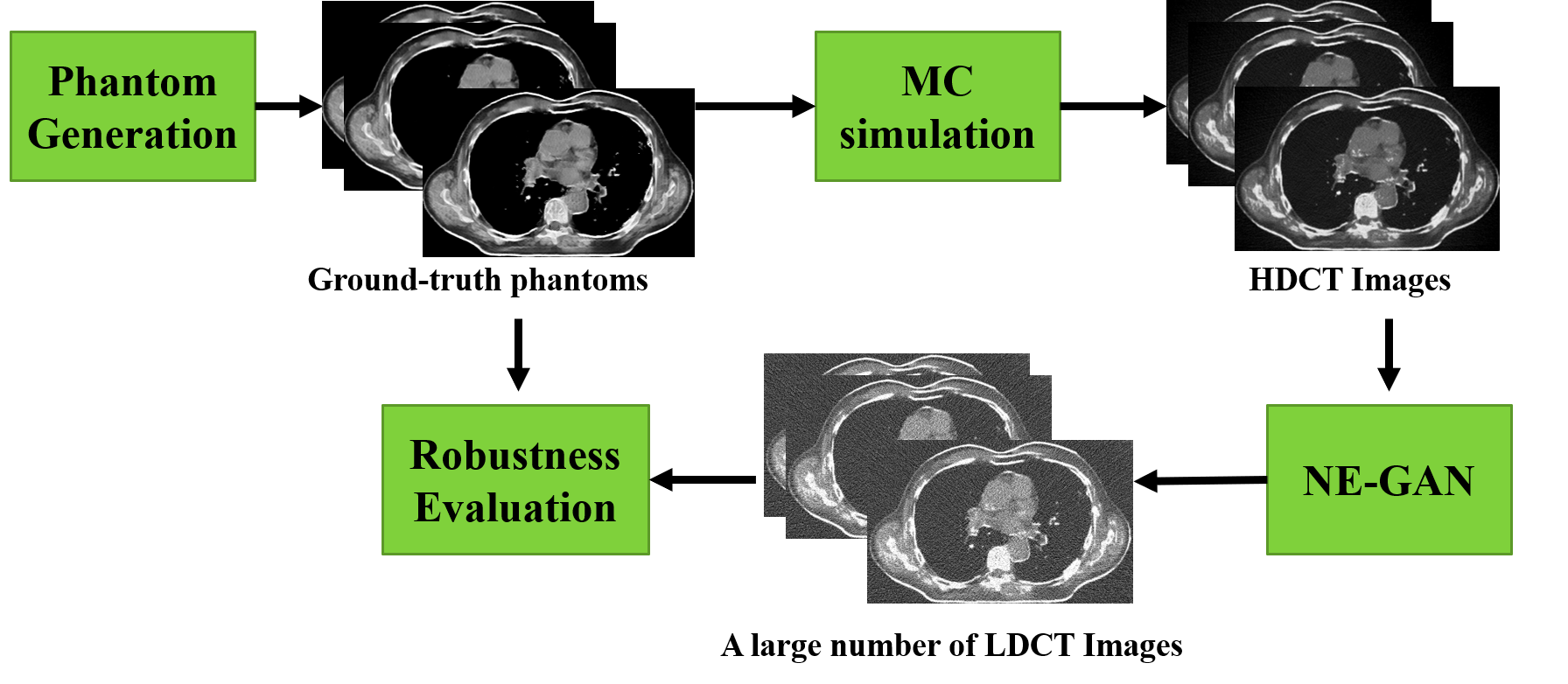}
    \caption{Virtual CT workflow for robustness evaluation of LDCT denoising algorithms.}
    \label{fig_flow}
\end{figure}

We presented a low-dose CT simulation method based on deep learning. Visual comparison and NPS-based noise property evaluation have demonstrated the effectiveness of the proposed method. One main advantage of the proposed NE-GAN is the high speed for simulation, which is extremely important when simulating a large number of LDCT images. For example, NE-GAN could be applied into a virtual CT workflow for robustness evaluation of LDCT denoising algorithms by generating a large number of LDCT images with the ground-truth, as shown in Fig. \ref{fig_flow}. Specifically, a CT generation method \cite{c25} is first used to generate many phantoms, which are forwarded to Monte Carlo (MC) simulation tools \cite{c26} to simulate the HDCT images. Then, NE-GAN can simulate a large number of LDCT images with a fast speed. Finally, the large number of LDCT images with the ground-truth can be used to test the robustness of the LDCT denoising algorithms.
In the future, we will further improve the simulation quality by adding some statistical constrains in the NE-GAN loss function, such as those based on the noise variance map and the noise power spectra.



\printbibliography

\end{document}